\documentstyle[prl,aps,multicol,epsf,amssymb]{revtex}
 
\newcommand{\be}{\begin{equation}}
\newcommand{\ee}{\end{equation}}           
\newcommand{\beas}{\begin{eqnarray*}}
\newcommand{\eeas}{\end{eqnarray*}}
\newcommand{\bea}{\begin{eqnarray}}
\newcommand{\eea}{\end{eqnarray}}

\begin{document}

\title{Kob-Andersen model: a non-standard mechanism for the glassy transition}

\author{S. Franz(*), R. Mulet(*)~\cite{MULET}, G. Parisi(**)}

\address{(*) The Abdus Salam Centre for Theoretical Physics, Condensed
Matter Group,\\
 Strada Costiera 11, P.O. Box I-34100, Trieste, Italy \\
(**) Universit\`a di Roma ``La Sapienza''\\
Piazzale A. Moro 2, 00185 Rome (Italy)  }

\date{\today}  

\maketitle 

\begin{abstract}
We present new results reflecting the analogies between the Kob-Andersen model
and other glassy systems. 
Studying the stability of the blocked configurations above and below
the transition we also
give arguments that supports their relevance for the glassy
behaviour of the model. 
 However we find, surprisingly, that the organization
of the phase space of the system is different from the well known 
organization of other mean-field spin glasses and structural glasses.

\end{abstract}

\pacs{PACS numbers: }

\begin{multicols}{2}
\narrowtext           
\section{Introduction}

In the last two decades our understanding of the glass transition in
spin systems has provided a coherent picture of glassy phenomena at
the mean field level\cite{BCKM}. These systems provide the simplest example of
systems where complex energy landscapes, with many local energy
stationary points, lead to vitrification and to the absence of
thermalization over large time scales.  At a mean field level, there
are spin glass models which show a dynamic transition very well
differentiated from the static one. Below the dynamical transition
temperature the system is unable to thermalize and get stacked at
energies above the equilibrium one. At the same value of the
temperature the space of equilibrium states gets disconnected.

The situation is different for real materials, where the sharp transition
is not present, and seems to be substituted by a smooth crossover to a
dynamics dominated by activated processes~\cite{Gotze}.  Progresses in the last
decade of research, however, indicate that many of the aspects of
the picture of the glass transition gained from the study of mean
field models may well apply to real materials.  In fact, while already
Kirkpatrick et al. ~\cite{Kirkpatrick,MePa} noticed the similarity between the thermodynamic
transition in structural glasses and spin glasses, later there were
found discontinuous transitions in spin glasses without quenched
disorder giving confidence that this analogy was not fortuitous
~\cite{JPBMM,mpr1}.  
Now, it has been recognized that at least in some approximation, classical
first principle models have a phenomenology which resembles to that
already found in mean field spin glasses ~\cite{BCKM,Frvi,CPV}. 

Moreover, an important feature of the phenomenology of glass forming
liquids is the formation of dynamical heterogeneities, characterized
by a spatial extension and a given life-time, both of which increase
as the temperature is decreased. These heterogeneities have been
found in numerical simulations of various realistic
models~\cite{dgp,kdppg} 
as well as 
in experiments on colloidal glasses~\cite{colloids}, and seem
to be a salient feature of landscape dominated glassy systems. In
order to quantify the presence of dynamic heterogeneity, it has been
proposed to study 4-point density dynamic correlation function. For
this quantity mean field theory predicts a peculiar time/temperature
dependence, in agreement with the one observed in numerical
simulations~\cite{FrPa}.

A related problem is the
physics of granular systems. Since, in general, these systems involve
many particles it is tempting to treat them using the methods of
statistical mechanics and, in fact, this approach has become an active
field of research for people with a background in this
field~\cite{HHL}. Particularly interesting have been the experimental results of
the group of Chicago who showed that a granular system subject to
continuous tapping develops a glass-like transition~\cite{Knight}.

An alternative view of the glass transition comes from kinetic models,
which have been useful to prove that complex energy landscapes are not
a necessary ingredient to have glassy phenomena.  Vitrification has
been found in simple lattice cellular automata models with kinetic
constraints. 

An interesting model belonging to this class is the Kob-Anderson
model~\cite{KoAn}, which has been shown by several studies to undergo a structural
arrest at densities larger than the threshold one. The phenomenology
of the model has been compared with qualitative agreement with the
Mode Coupling Theory~\cite{Gotze}, which proposes an intrinsically kinetic view of
the glass transition.

Despite the fact that the mechanisms leading to structural arrest in the
model can not be related to a non-trivial energy landscape, many aspects
of the physics of the model are very similar to the phenomenology of
mean field glassy systems. In particular the blockage to the threshold
density found in aging experiments in \cite{Se,levin} closely reminds the
blockage of the internal energy in the dynamics of mean field spin
glasses. In addition, it has been recently shown that during aging
dynamics, the system probes portions of phase space close to generic
random blocked configurations \cite{bklm}, again, in a way reminiscent of
mean field glassy phenomenology. 

It is tempting in such situation to invoke some universal mechanism
underlying structural arrest both in landscape dominated systems, and
in kinetically constrained ones.  In this paper we investigate the
limits to which these analogies can be pushed and to what extent the
mechanisms are really universal.

The Kob-Andersen model, that first emerged as a  model to describe the
dynamic transition in structural glasses, has also been recently used as a
simple model for the jamming transition in granular 
systems~\cite{SeAr,Se,bklm}. While
many other lattice models have been proposed to study similar
problems ~\cite{prltetris,nicodemi}, this one has the advantage of a trivial
equilibrium state that can be easily characterized analytically without
further assumptions, and that therefore allows to a clear comparison
between the static and the dynamic properties of the system.

A first test to which we submit the KA model is the computation of 
the dynamical susceptibility associated to a 4-point function, and
show that it behaves as expected from mean field theory. 

A second point that we investigate is the nature of the blocked
states. A lot of emphasis has been recently made about the role of
stationary points of the energy surface in glassy dynamics, and the
ideal mode coupling transition has been associated to the point where
the number of negative directions of the inherent saddles 
vanishes~\cite{Sciortino,Frvi}.

In other words, while above, the MC point the dynamics spends a lot of
time close to saddles, below, it spends its time close to minima. 
The work of \cite{bklm} has emphasized the role of the blocked states
during aging in the KA model. In this paper we show that while this
states are responsible of the glass-like properties of the model,
their geometrical organization differs from the organization of the
states found in spin and structural glasses models.

The remaining of the paper is organized as follows: In section
\ref{sec:mod} we define the model and discuss the current
understanding in the literature. In section \ref{sec:res} we present
our numerical simulations and discuss the results, finally in
\ref{sec:conc} the conclusions are outlined.
 
\section{The model}
\label{sec:mod}

The Kob-Andersen model is a lattice gas defined in a three dimensional
lattice of size $L^3$ where $N$ particles interact by hard core
repulsion (i.e only one particle can be in a site at a given moment)
and that evolves using dynamic rules (described below) which are
symmetric in time, so that the detailed balance is satisfied.

We first choose a particle at random, then the particle moves if the
following conditions are satisfied

\begin{itemize}
\item the neighboring site is empty
\item the particle has less than 4 nearest neighbors
\item the particle will have less than 4 nearest neighbors after it
      has moved
\end{itemize}

Already in the first paper devoted to the model, it was shown that the
dynamics becomes slower when the density $\rho$ in the system
increases. In particular the diffusion coefficient vanishes as a power
low at the critical density $\rho_c \approx 0.88$ independently of the
system size ruling out the possibility that this phenomena could be
associated with the formation of an infinite backbone in the system~\cite{KoAn}.

More recently, the system has been studied in the presence of a
chemical potential~\cite{bklm}. Despite of the trivial equilibrium properties of
the system, it was proved in a series of papers that: Below a certain
critical density ($\rho_c \approx 0.88$) the system is not ergodic,
independently of the chemical potential the density gets blocked at a
density lower that the critical density. This density depends on the
''cooling rate''.  Detailed linear response numerical experiment have
shown the emergence of an effective temperature. This effective
temperature coincides very well with the ''Edwards'' temperature,
which can be computed from a uniform measure over blocked
configurations. The spatial correlation function obtained by a slow
annealing close to $\rho_c$ is very similar to the one expected from
the Edwards' measure.

While these results and other presented below closely resembles the
macroscopic behavior of many glassy systems we will probe that this
analogy can hardly be extended to the organization of the phase space
of the system and therefore suggests a non-universality of the glassy
mechanism.

\section{Results and Discussion}
\label{sec:res}

\subsection{The non-linear susceptibility}

It has been recently stressed that glassy dynamics has an
inhomogeneous character. There one observes regions of correlated
motion whose sizes depend of a control
parameter, and that survive for a characteristic time also
parameter dependent~\cite{FrPa}. This character can be revealed studying the
4-point density correlation function. In lattice gas models this can
be defined as

\begin{equation}
\chi(t)=N(\langle q^2(t)\rangle -\langle q(t)\rangle^2)
\end{equation}

\noindent where 

\begin{equation}
q(t)=\frac{1}{N \rho(1-\rho)}\sum_i (n_i(t)n_i(0)-\rho^2).
\end{equation}

\noindent and the average is performed with respect to different samples. 
In \cite{dfgp}, based on the theory of disordered systems, it was
proposed that in the deep supercooled phase, due to the inhomogeneous
character of the dynamics, $\chi(t)$ should display a maximum as a
function of time. This maximum appears
 more and more pronounced, and displaced to larger
and larger times as the temperature is decreased (which is equivalent
in the KA model to increase the density of particles in the system). 

This behavior reflects
the geometrical structure of the phase space visited during the evolution
of the system and the physics of time scale separation. Along its
evolution the system remains blocked in regions of phase space that
have longer and longer time life, and with more and more correlated
density fluctuations as the density is increased. These regions lye
close to saddle points of the energy surface, which have lower and
lower number of escaping directions, until at the dynamical transition
the number of negative directions tend to zero.

\begin{figure}
\epsfxsize=0.95\columnwidth 
\epsffile{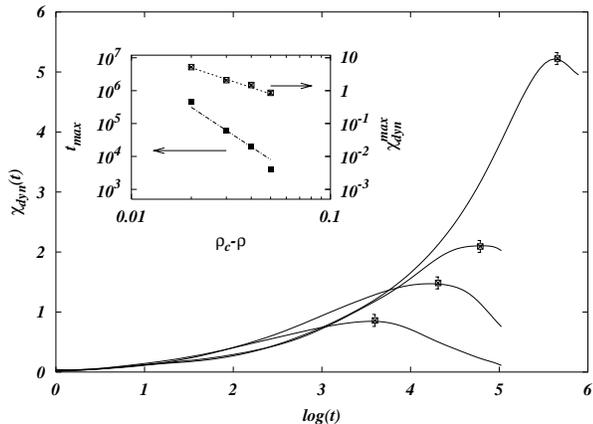}

\caption[0]{The non linear susceptibility in the KA model as a function
of time for different densities. From top to bottom $\rho=0.86$,
$0.85$, $0.84$, $0.83$. The inset shows $t_{max}$ and $\chi_{dyn}^{max}$
close to the transition, the lines are power laws guide to the eyes 
with exponents $4$ and $2$ for $t_{max}$ and  $\chi_{dyn}^{max}$ respectively}
\label{fig1}
\end{figure}

We have investigated the behavior of the function $\chi(t)$ as a
function of the density in the KA model, simulating systems with size 
$L= 10$ and $L=20$, and averaging over 100 samples. 
In figure (\ref{fig1}) we show the behavior of the non linear
susceptibility as a function of time for various values of the
density, showing that qualitatively this function behaves in the 
same way found in simulations of liquids and in spin
models~\cite{glotzer} and theorized in~\cite{dfgp}. 
The maximum, as well as the position of the maximum seem to diverge 
at the dynamical critical density as power laws with exponents
4 and 2 respectively. 

It is interesting to note that in the low density phase, although the
KA probes the configuration space ergodically, the dynamics develops
correlations which are similar to the ones seen in structural
glasses. Even if the infinity time limit of the susceptibility is
trivial, due to the fact that the system visit the space ergodically,
the dynamics is highly correlated. 

In ref. \cite{bklm} it was emphasized that during the aging dynamics 
of the model, the measure concentrates on the blocked configurations. 
It remains to be explained why in the aging experiments the system 
blocks exactly at the value of the density where the fixed density
experiments are not ergodic. 

\subsection{The stability of the blocked states}

In order to understand the role of the blocked configurations as a
function of the density, we studied the stability of the blocked
states with respect to the move of a single particle, in a way
analogous to study of the local stability in an energy landscape. 

To this end, we used the auxiliary model already
introduced in~\cite{bklm}. There, 
one associates to each configuration of the lattice gas an energy equal
to the number of particle which are free to move according to the KA
dynamics.  The thermodynamics of this auxiliary model can be studied
easily numerically as a function of an auxiliary temperature
$T_{aux}=1/\beta_{aux}$. For example, the Edwards entropy as a function of
the density can be obtained from the thermodynamic integration of:

\begin{equation}
S(\rho)=H(\rho)+\int_0^\infty d\beta_{aux}\;
\beta_{aux}U_{aux}(\beta_{aux}). 
\end{equation}

\noindent where $H(\rho)=-\rho \log(\rho)-(1-\rho) \log(1-\rho)$
is the entropy of the model at $\beta_{aux}=0$.

With the help of this auxiliary model 
we generated blocked configurations at different densities 
by lowering the auxiliary temperature slowly, in
order to reach a typical Edwards state at $T_{aux}=0$.  We then moved a
particle to an arbitrary site, and let the system evolve according to the KA
dynamics. The perturbation to the blocked state causes some particle 
to become free to move, so we detected the number of particles that as 
a consequence of the perturbations move at least once. 
We observed two kind of events: 

\begin{itemize}
\item
either the displacement of
a particle has a little effect, and after a few particle move the
system evolves again to a blocked state.

\item or the displacement of the particle 
completely unblocks the system which starts to move ergodically, after
a time which depends on the particle density.
\end{itemize}

\begin{figure}
\epsfxsize=0.95\columnwidth
\epsffile{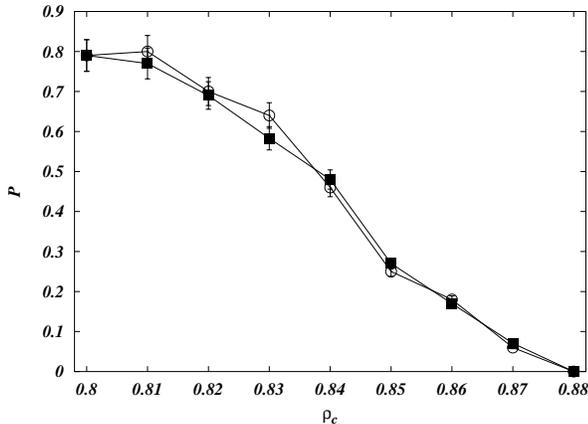}

\caption[0]{The unblocking probability as a function of the density.
$\circ$ Data from the block configurations obtained with the auxliary
model, $\blacksquare$ Data from block configurations obtained taking a
random configuration}
\label{fig2}
\end{figure}

In figure \ref{fig2} we display the estimated unblocking
 probability, obtained generating 1000 blocked
configurations using the simulated annealing as describe above 
 and checking how many unblock, as a
function of the density (white circles in figure 2).
At low density, this probability is different to zero
in the low density phase which means that
even if the system approaches by chance a
blocked state, it will always depart from it.  The value of the density
at which the unblocking probability vanishes is compatible with the
critical density $\rho_c=0.88$ of the dynamical instability. 

We have also checked that configurations generated with a quenched of
the auxiliary model form $T_{aux}=\infty$ to $T_{aux}=0$ give rise to
the same unblocking probability, hinting that  all blocked
configurations have the same stability properties. 

\subsection{The geometrical structure of the blocked states}

As we have seen the blocked states are unstable for densities below
threshold, while they are stable above threshold. One can ask if this
stability property of the blocked states is in correspondence with
their geometrical organization in space as it happens in mean field
spin glasses \cite{cavagna}. There the stationary points of the energy are
saddles above threshold and minima below.  It has been shown that the
minima are disjoint below threshold: the maximal overlap between
minima $q_0$ is smaller than one, and tends to one as the energy tends
to the threshold energy from below, meaning that at that value of the
energy one finds minima arbitrarily close from a given minimum.  Above
the threshold, there is no gap between saddles, and indeed one can
show that the logarithm of the number of saddles at overlap $q$ from a
given saddle grows as $\Sigma(E,q)\approx N(1-q)$. The threshold
energy appears as a kind of percolation threshold for stationary point
in the configuration space. It is worth noticing that while 
this organization is at the heart of the mechanism of dynamical
transition, there is not sign in the Gibbs measure that reveals this
structure. 

One might wonder if such an organization of stationary points is also
found in the KA model. In order to investigate this
point we have studied the number of blocked states at fixed overlap
with a generic blocked states introducing in the auxiliary model a
term $N\frac{\epsilon }{\beta_{aux}}(q(\{n_i\},\{n_i^0\})-q)^2$, which
for large $\epsilon $ constrains the configurations to have the wanted
overlap with the reference configuration. In the simulations the value
$\epsilon =50$ proved to be effective to get the wanted values of the
overlap for all densities. We performed a simulated annealing of the
model and we extracted the zero temperature entropy from thermodynamic
integration of eq(1).

\begin{figure}
\epsfxsize=0.95\columnwidth
\epsffile{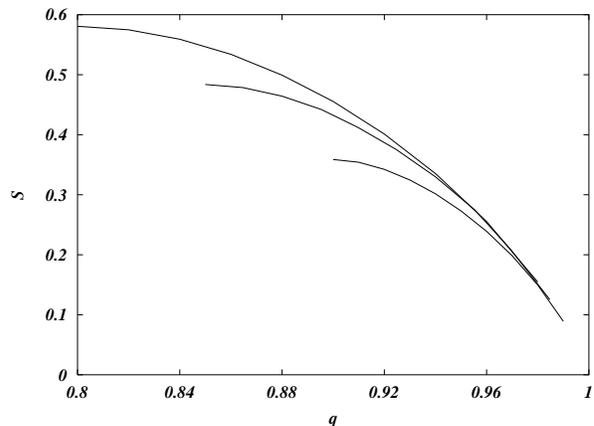}

\caption[0]{The entropy of the blocked states as a function of $q$ for 
densities (from top to bottom) $\rho=0.8,0.85,0.9$.}
\label{fig3}
\end{figure}

In figure \ref{fig3} we show the entropy as a function of 
the overlap for different values of the density. It is apparent from the 
figure, that both above and below the transition, the curves behave
qualitatively in the same way, with an entropy as a function of $q$ 
which departs from zero regularly at $q=1$. This 
indicates that the structure of the blocked states does not change as we
cross the transition probing that 
the dynamic transition in this model is not linked to the
geometrical organization of the blocked states. 

\section{Conclusions}
\label{sec:conc}

In this paper we have investigated the mechanism which lead to
vitrification in kinetic lattice gases, in comparison with the 
one responsible of glassy phenomena in mean field models with 
random Hamiltonian. 

We have found that in the low density phase, 
despite the trivial character of density correlation implied by 
the equilibrium distribution, as it happens in supercooled liquids
the dynamics has an heterogeneous character. The four point correlation
function displays a maximum as a function of time that grows as a 
power of $\rho-\rho_c$, and the dynamics remain correlated for times 
which also grow as a power of the same quantity.

We have shown that 
the blocked states are unstable and play no role
 in the low density phase, while they
are stable and act as attractor in the high density phase. 
Furthermore, we have investigated if this stability property 
corresponds to a
different organization of the blocked states in phase space, with
negative result. In both dynamical phases (low and high density), 
one finds blocked states at 
all distances from any blocked state. 

Then, we can conclude that despite the similarities in the dynamics, 
the glassy phenomena in kinetic models, at least at the level 
of the phase space organization, are very different from the ones
found in systems with a complex energy landscape. 

\section*{Acknowledgments}
 We thanks J. Berg and M. Sellito for many useful discussions and
 suggestions.

\end{multicols}  

\end{document}